\documentclass[aps,prl,reprint,amsmath,amssymb,superscriptaddress,longbibliography]{revtex4-2}

\pdfoutput=1

\newcommand{\RCMO}{Rb$_2$Cu$_2$Mo$_3$O$_{12}$\xspace}
\newcommand{\CCMO}{Cs$_2$Cu$_2$Mo$_3$O$_{12}$\xspace}

\usepackage{graphicx}
\usepackage{dcolumn}
\usepackage{bm}
\usepackage{amsmath}
\usepackage{amssymb}
\usepackage{xcolor}
\usepackage{xspace}
\usepackage{xfrac}
\usepackage{color,soul}
\usepackage{graphicx}
\usepackage{hyperref}
\hypersetup{colorlinks=true,linkcolor=blue,citecolor=blue,urlcolor=blue}
\usepackage{soul}

\begin{document}
	
	\title{Dielectric Relaxation by Quantum Critical Magnons}
	
	\author{Daniel Flavi\'{a}n}
	\affiliation{Laboratory for Solid State Physics, ETH Z{\"u}rich, 8093 Z{\"u}rich, Switzerland}
	\author{Pavel A. Volkov}
	\affiliation{Department of Physics, Harvard University, Cambridge, Massachusetts 02138, USA}
	\affiliation{Department of Physics, University of Connecticut, Storrs, Connecticut 06269, USA}
	\affiliation{Department of Physics and Astronomy, Center for Materials Theory, Rutgers University, Piscataway, NJ 08854, USA}
	\author{S. Hayashida}
	\affiliation{Laboratory for Solid State Physics, ETH Z{\"u}rich, 8093 Z{\"u}rich, Switzerland}
	\affiliation{Present address: Max-Planck-Institut f\"ur Festk\"orperforschung, Heisenbergstraße 1, 70569 Stuttgart, Germany}
\author{K.~Yu.~Povarov}
\affiliation{Laboratory for Solid State Physics, ETH Z{\"u}rich, 8093 Z{\"u}rich, Switzerland}
	\affiliation{Present address: Dresden High Magnetic Field Laboratory (HLD-EMFL) and W\"urzburg-Dresden Cluster of Excellence ct.qmat, Helmholtz-Zentrum Dresden-Rossendorf, 01328 Dresden, Germany}	
	\author{S. Gvasaliya}
\affiliation{Laboratory for Solid State Physics, ETH Z{\"u}rich, 8093 Z{\"u}rich, Switzerland}
	\author{Premala Chandra}
	\affiliation{Department of Physics and Astronomy, Center for Materials Theory, Rutgers University, Piscataway, NJ 08854, USA}
	\author{A. Zheludev}
	\affiliation{Laboratory for Solid State Physics, ETH Z{\"u}rich, 8093 Z{\"u}rich, Switzerland}
	
	\begin{abstract}
We report the experimental observation of dielectric relaxation by quantum critical magnons. Complex capacitance measurements reveal a dissipative feature with a temperature-dependent amplitude due to low-energy lattice excitations and an activation behavior of the relaxation time. The activation energy softens close to a field-tuned magnetic quantum critical point at $H=H_c$ and follows single-magnon energy for $H>H_c$, showing its magnetic origin. Our study demonstrates the electrical activity of coupled low-energy spin and lattice excitations, an example of quantum multiferroic behavior.
 
	\end{abstract}
	\maketitle

Magnetic insulators have long served as experimental prototypes for fundamental studies of thermal and quantum phase transitions. These materials can often be carefully tuned to quantum critical points (QCPs), 
for example by   an applied magnetic field. In this category are soft-magnon and saturation transitions in antiferromagnets (AFs), often described in terms of a Bose-Einstein condensation (BEC) of magnons \cite{Batyev1984,Giamarchi1999,Nikuni2000,Giamarchi2008,Zapf2014}. In  some real materials, magnetoelectric coupling results in dielectric anomalies at these magnetic-field-controlled transitions \cite{Kim2014,Schrettle2013,Povarov2015,Kimura2016,Kimura2017,hayashida2021}. In most known examples, the electric polarization is merely a passive participant, reflecting the evolution of spin correlations; the latter is adequately described by purely magnetic  microscopic Hamiltonians. There is no reason for this to always be the case. The search for qualitatively new dielectric phenomena at magnetic QCPs and the so-called ``multiferroic'' quantum critical regime \cite{Narayan2019} continues.

In this Letter we present one such phenomenon, namely dipolar relaxation induced by magnons
tuned through a magnetic-field-controlled QCP in the quantum spin system \CCMO. This relaxation manifests itself as a frequency-dependent dielectric anomaly with a characteristic temperature scale distinct from both the classical transition temperature and the magnon Zeeman gap. We show that the anomaly is well described by a relaxation model with a field-independent background dielectric constant attributed to lattice degrees of freedom alone. However, the relaxation time has an activated behavior with an {\em  energy barrier closely following the single-magnon energy}. We conclude that the observed anomaly arises as a result of the interaction between low-energy lattice and quantum-critical spin degrees of freedom.

Our subject of study, \CCMO, has previously attracted attention as a  frustrated ferro-antiferro $S=1/2$ quantum spin chain system ~\cite{Hase2005,Fujimura2016}. It is structurally similar \footnote{The exact crystal structure of \CCMO was determined by single crystal X-ray diffraction in this work and is reported in the Supplementary Material} to the more extensively studied \RCMO ~\cite{Hase2004,Hase2005,Yasui2014,Hayashida2019,hayashida2021,Ueda2020}. The Cs-based compound orders magnetically in three dimensions at $T_{\rm N} = 1.85$~K~\cite{Fujimura2016,Flavian2020}. In applied magnetic fields it has a rather complex phase diagram, and achieves saturation at $\mu_0H_c\sim 7.7$~T via a magnon-BEC-type QCP \cite{Flavian2020}. Inspired by the fact that the magnetic field-induced AF phase in \RCMO is also  ferroelectric (FE) \cite{Reynolds2019,Ueda2020,hayashida2021}, we performed detailed dielectric permittivity studies of \CCMO as a function of temperature ($T$), magnetic field ($H$) and probing frequency of the electric field ($\omega$). To this end, we deposited conductive electrodes on opposing facets of a single crystal sample, such that the electric field was $\mathbf{E} \parallel \mathbf{c}^\ast$. The data were collected at dilution refrigerator temperatures in a capacitance bridge setup similar to that used in Ref.~\cite{hayashida2021}. The change in complex capacitance was  monitored as the external parameters and measurement frequency were varied. Experimental details are given in the Supplementary Material.

	\begin{figure*}[tbp]
		\centering
		\includegraphics[scale=1]{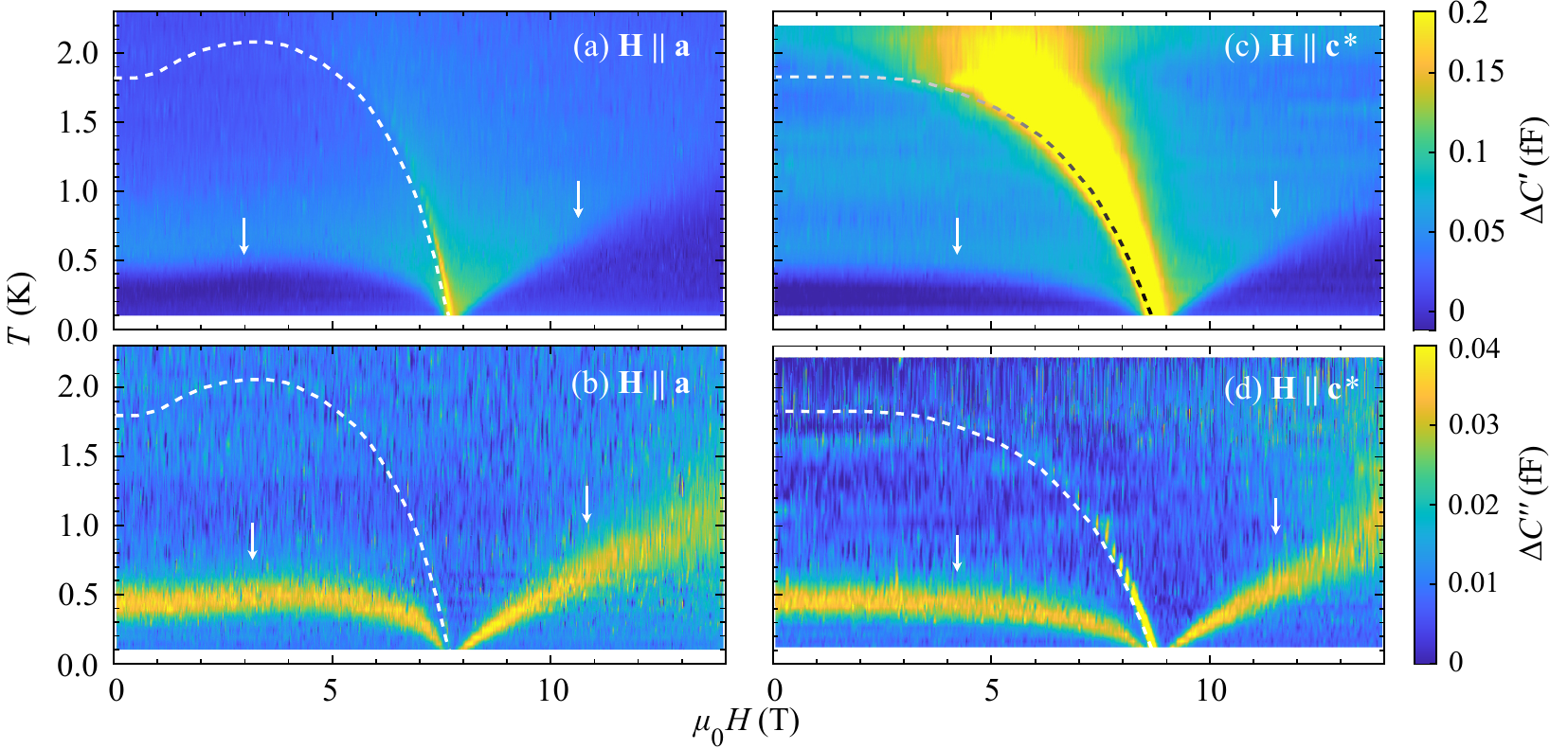}
		\caption{False color plots of complex capacitance $\Delta C = \Delta C^{\prime} + i \Delta C^{\prime\prime}$. The electric field is applied parallel to the $\mathbf{c}^*$ axis. The magnetic field is applied along two orthogonal crystallographic directions: (a,b) $\mathbf{H}\parallel\mathbf{a}$ and (c,d) $\mathbf{H}\parallel\mathbf{c}^*$. The color scale is shared across (a)/(c), and (b)/(d), respectively. It has been chosen to highlight the novel low temperature anomalies (arrows). As a result some stronger peaks at the magnetic phase boundary (dashed line, Ref.\cite{Flavian2020}) are blown out. }
		\label{fig:overview}
	\end{figure*}

Raw capacitance data collected at $\omega=1$ kHz are presented in Fig.~\ref{fig:overview}. The real ($\Delta C'$) and imaginary ($\Delta C''$) components, proportional to the respective components of the dielectric constant  $\varepsilon'$ and $\varepsilon''$, 
are displayed in false colors as a function of temperature and magnetic field. 
	At precisely the phase boundary separating the magnetically ordered and paramagnetic states (dashed line), $\Delta C'$ shows a sharp narrow peak, indicating the divergence of $\varepsilon'$. The magnitude of the anomaly is markedly different for $H\parallel a$ and $H\parallel c^*$ \footnote{See the Supplementary Material for further details.}. This feature is very similar to that previously seen in \RCMO \cite{hayashida2021}. The only obvious difference is that the Cs compound orders already in zero applied field, while its Rb-based counterpart is a quantum paramagnet and only shows spontaneous AF order when magnetized  \cite{Hayashida2019}. 
	
	\begin{figure}[tbp]
		\centering
		\includegraphics[scale=1]{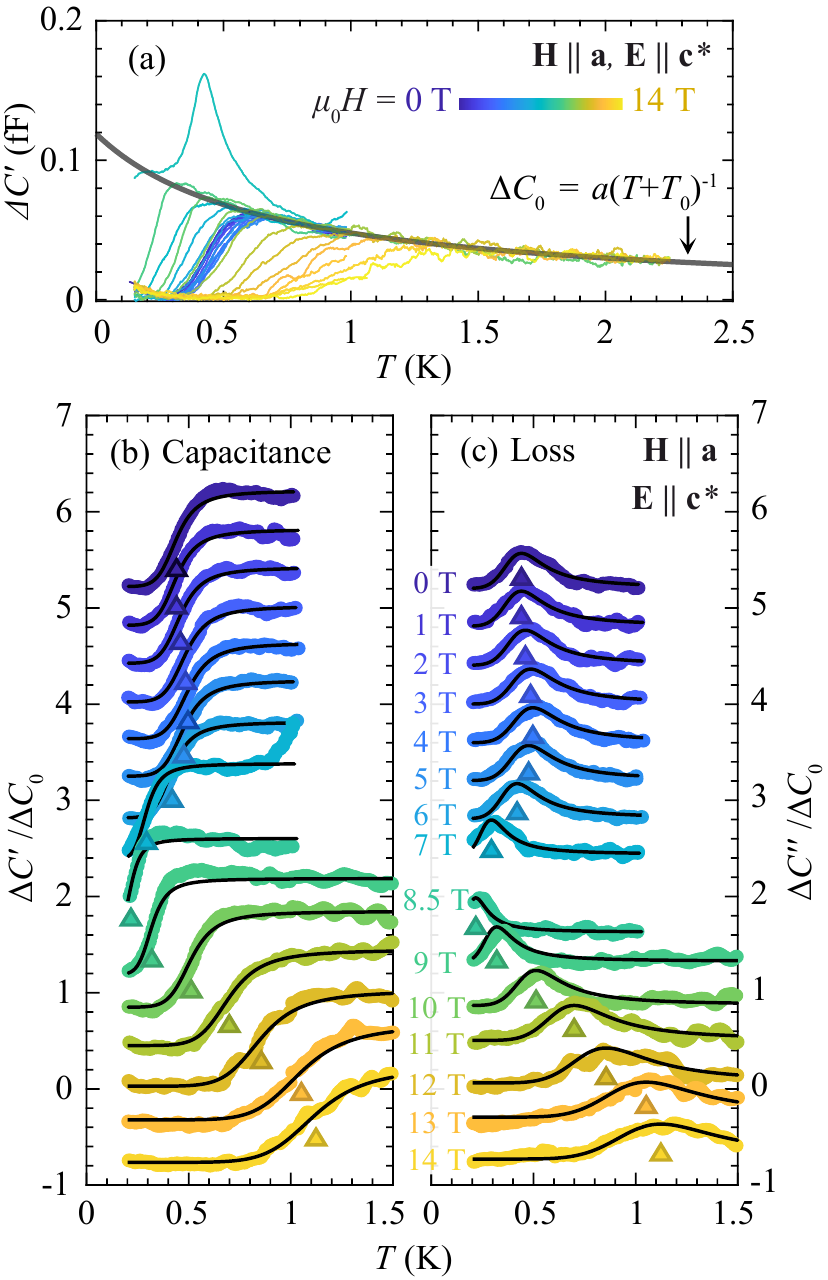}
		\caption{Constant-field measurements of complex capacitance. (a) Regardless of the applied field, all capacitance curves collapse on a single line (solid black line) at elevated temperatures.  This data shown in (b)- real part and (c)- imaginary part were normalized by the thus determined ``background''. The geometry corresponds to that of Fig.~\ref{fig:overview}.a,b: $\textit{E}\parallel\textit{c}^*$ and $\textit{H}\parallel\textit{a}$. In all panels,  same colors correspond to same magnetic field values. An offset of 0.4 units has been added between sets for visibility. Black solid lines show the best fit to the Cole-Cole model of Eq.\ref{eq:ColeCole}. Note that both panels show the same scale and units. Triangles ($\blacktriangle$) denote the position of $T^*$. }
		\label{fig:fits}
	\end{figure}	
	
In both materials, this peak represents the onset of ferroelectricity simultaneous with magnetism in the ordered phase. Divergence of the capacitance implies that, along with any magnetic order, polarization is itself  a primary order parameter of the transition, and is critical at the QCP at $H_c$ \cite{hayashida2021}.  This behavior is explained by the spin-current or inverse Dzyaloshinskii-Moriya mechanism of magneto-electric coupling \cite{Katsura2005,Mostovoy2006,Tokura_2014,Kimura2016}. In this model $P\propto M_\perp \cdot M_\parallel$, the product of transverse staggered ($M_\perp$) and longitudinal uniform ($M_\parallel$) magnetization,  respectively \cite{hayashida2021}. The former is the symmetry-breaking magnetic order parameter of the AF phase, while the latter is merely induced by the applied magnetic field. Consistently, no dielectric anomaly is found at the phase boundary at zero applied field, where no uniform magnetization is present. The strength of the anomaly increases as an external magnetic field progressively magnetizes the system.
	
\begin{figure}
		\centering
		\includegraphics[width=\columnwidth]{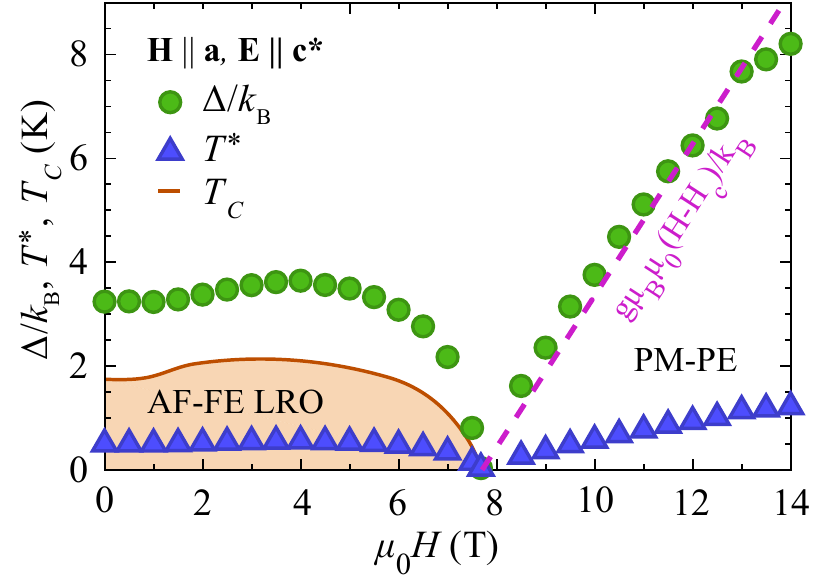}
		\caption{Characteristic energy barrier extracted from Cole-Cole model (cirlces). Above the QCP, it follows the Zeeman energy of a single magnon, given by a pink dashed line. Triangles represent the position of $T^\ast$, at which the relaxing dissipative anomaly is observed. For completeness, the phase boundary is depicted, showing the PM-PE (paramagnetic-paraelectric) and AF-FE phases (shadowed) \cite{Flavian2020}. }
		\label{fig:phdiag}
	\end{figure}

The central new observation of the present study is an additional feature in the dielectric permittivity, one that has no analog in \RCMO. In \CCMO it shows up as a broad peak-like feature in $\Delta C''(T)$ at temperatures well below $T_c$ in the ordered phase. At fields above saturation it appears at a temperature that increases with applied field.
It is accompanied by a rapid step-like change of $\Delta C'(T)$ (see Fig.~\ref{fig:fits} for details). Unlike the divergence at the phase boundary, these additional features show virtually no difference in magnitude between measurements under different magnetic field orientations (Fig.~\ref{fig:overview}). Moreover, their magnitude is not particularly field-dependent, and they persist even in the absence of applied field. Obviously, their origin must be entirely different. Yet, both effects must be intimately linked, as they both coalesce and become critical at the QCP. We henceforth refer as $T^*$ to the temperature at which $\Delta C''$ peaks. The values of $T^*$ at different fields are displayed in Fig.~\ref{fig:fits} (triangles) and Fig.~\ref{fig:phdiag}.

	Several simple scenarios for the $T^*$ feature can be ruled out. First, the feature must be a bulk-, rather than a surface- or finite-size-effect, since a comparable anomaly at the same $T^\ast$ is also observed in polycrystalline samples \cite{Note2}. Second, there is no thermodynamic phase transition at these temperatures \cite{Flavian2020}. Such a large dissipation is not something that one would expect at a thermodynamic phase transition. A glass-like freezing transition is also unlikely, due to the absence of any hysteresis or history-dependence in our measurements. 
	
	Step-like behavior of $\varepsilon'$ above a magnetic field-induced saturation QCP has been reported previously in another quantum magnet  Ba$_2$CoGe$_2$O$_7$ \cite{Kim2014}. There it was attributed to the fluctuations of the AF order, the characteristic energy/ temperature of the anomaly coinciding with that of the field-induced single-magnon Zeeman gap $g \mu_B \mu_0(H-H_c)$. This interpretation can not be carried over to \CCMO. In our material,  above $H_c$  the scale of $T^*$  is roughly {\em ten times smaller} than the Zeeman gap (triangles vs. the dashed line in Fig.~\ref{fig:phdiag}). A different mechanism must be at play.
	
	Since the data shown were collected at finite frequency of 1~kHz, one could suspect a proximity of a mechanical resonance in the measurement setup that couples to our probe. Frequency-dependent measurements firmly rule that out \cite{Note2}: while some mechanical resonances are indeed detected, none are found in the  immediate neighborhood of 1~kHz. Also, all appear to be excited by Lorentz forces in the connecting wires, and therefore vanish in the absence of an applied magnetic field, very much unlike the feature under discussion. The final suspicion is that slow dielectric relaxation may arise from trapping of charge carriers \cite{Scott2015}. In our case this seems unlikely, as \CCMO is a good insulator and is transparent in bulk at room temperature.

		The origin of the $T^*$ feature is revealed in a quantitative analysis of the data. Noting the strong dissipative component, we attempted to fit the measurements with the phenomenological Cole-Cole model for dielectric relaxation \cite{Cole-Cole,kremer2002broadband}:
	\begin{equation}
		\Delta C  = \frac{\Delta C_0(T)}{1+(i\omega\tau(T))^{1-\alpha}}.
		\label{eq:ColeCole}
	\end{equation}
 Here $\Delta C_0(T)$ is proportional to the static dielectric constant $\varepsilon(T)$ and $\tau(T)$ is the relaxation time. The exponent $\alpha=0$ corresponds to the Debye relaxation model with a single relaxation time \cite{debye1929polar}, while $\alpha<1$ accounts for a distribution of relaxation times within the solid.

	Qualitatively, relaxation may be expected to occur faster at high temperature, and $\tau(T)$ is expected to decrease with increasing temperature. Therefore, at high enough $T$,  $\Delta C$ \eqref{eq:ColeCole} should be proportional
 to the static dielectric constant. In Fig. ~\ref{fig:fits}(a), we demonstrate that most of the measured $\Delta C'$ curves {\em collapse on a single, field-independent, line at} $T \gtrsim T^*$. The only exception is the data close to the QCP, where the sharp dielectric anomaly at the transition overwhelms the background. The resulting universal curve $\Delta C'_0(T)$ is well-fitted by the form $a(T+T_0)^{-1}$ ($T_0=0.68(1)$~K,  Fig. ~\ref{fig:fits}(a)). The lack of a field-dependence of $\Delta C'_0(T)$ implies that this contribution is unrelated to magnetism and is entirely due to the crystal lattice. One could suspect the origin to be a  yet undetected nearby structural transition, such as the one seen at around 20~K in \RCMO \cite{hayashida2021}. On the other hand, a lattice-driven ferroelectric transition is unlikely, due to the modest absolute value of the dielectric constant in \CCMO ($\epsilon \approx 5.5$, see Supplement). Moreover, the relative change in the dielectric constant due to the temperature-dependent background is smaller than $1\%$ \cite{Note2}. We conclude that $\Delta C'_0(T)$ is due to some yet to be identified  {\em persistent low-energy electric dipole degrees of freedom in} \CCMO.

	With $\Delta C'_0(T)$ characterized, the parameter $\alpha$ in \eqref{eq:ColeCole} can be determined from the plots of the real vs. imaginary parts of $\Delta C$ (Cole-Cole plots). As discussed in the Supplement, the data {\em at all fields and temperatures} are well described by {\em a single value} $\alpha=0.2$. This supports  \eqref{eq:ColeCole} being an appropriate description. The only remaining fit parameter not determined globally is the characteristic time scale  $\tau(T,H)$.
	
	Except for the immediate proximity of QCP, at each value of magnetic field, excellent agreement with experiment is obtained assuming an activated behavior
	\begin{equation}
		\tau(T,H) = \tau_0 \exp[\Delta(H)/k_\mathrm{B}T].
	\end{equation}
	In this expression $\Delta(H)$ is the ``barrier height'' and $\tau_0$ the ``attempt time''.
	The resulting fits are shown in Fig. ~\ref{fig:fits}. A {\em single field- and temperature- independent} $\tau_0\approx 10^{-6}$~s was found to be sufficient to capture the data quantitatively. We have also checked that other values of $\tau_0$ reduce the fit quality \cite{Note2}. Note that {\em the real and imaginary parts are reproduced simultaneously} using the same values of $\Delta(H)$ for all temperatures. In contrast, if we were dealing with glassy behavior, relaxation times would be expected to show a divergent behavior at a finite temperature \cite{glinchuk1990,kremer2002broadband}. This is yet another argument against strong disorder effects and glassiness, despite the distribution of relaxation scales indicated by $\alpha=0.2$.
	
\begin{figure}
		\centering
		\includegraphics[scale=1]{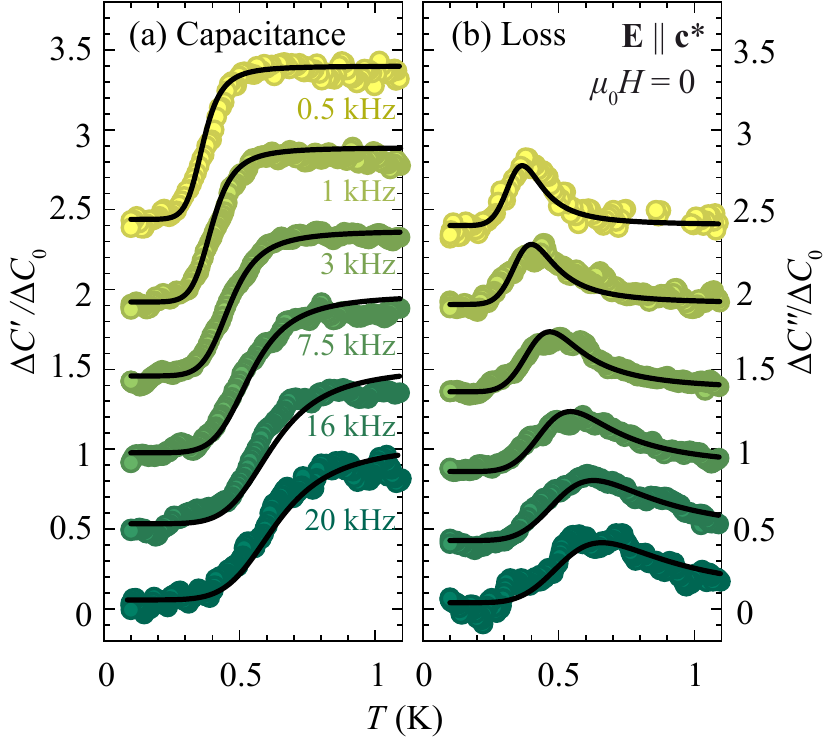}
		\caption{Frequency dependence of complex capacitance
			at zero magnetic field, (a)-real and (b)-imaginary parts. The excitation frequency is consistently color-coded in the two panels. An offset of 0.5 units has been added for clarity. Black lines show the best fit to Eq.\ref{eq:ColeCole} with parameters fixed from Fig.~\ref{fig:fits}. }
		\label{fig:freq}
	\end{figure}

	Equation \eqref{eq:ColeCole} implies a very specific form of frequency dependence. In Fig.~\ref{fig:freq} we demonstrate that the parameters extracted from the 1~kHz data also describe the capacitance measured at other frequencies ranging over more than a decade {\em without any additional fitting}. The conclusion is that the  relaxation time $\tau(T,H)$ is {\em independent of probing frequency and is an intrinsic property of} \CCMO.
	
	The combined results of our Cole-Cole analysis are borne out in Fig.~\ref{fig:phdiag}. The activation energy $\Delta(H)/k_\mathrm{B}$ is shown alongside with the temperature of long range ordering $T_c$ \cite{Flavian2020} and $T^\ast$. Strikingly, for $H>H_c$, $\Delta(H)$ coincides with the single-magnon Zeeman gap $g \mu_B\mu_0 (H-H_c)$, with $g=2.16$ \cite{Hayashida2019, Ueda2020}. This strongly suggests that the $T^\ast$ feature arises from the relaxation of low-energy structural dipole moments (the ones responsible for the background dielectric constant discussed above) by single magnons. The isotropy of $T^\ast$ with respect to the field orientation naturally follows from the very isotropic $g$-factor and saturation magnetization \cite{Flavian2020}. In the ordered phase, $T^\ast$ does not change much at low fields, but goes to zero as $H\rightarrow H_c$. This behavior is also consistent with a single-magnon process within the ordered phase.
	
	To put our findings in perspective, exponentially activated relaxation times are often observed in ferro- and para- electrics \cite{bidault1994,viana1994,ang1999}. In those cases, the activation energy usually corresponds to that of of a longitudinal optical phonon \cite{scott_1999} - an excitation carrying a finite dipole moment. In our case though, the dielectric relaxation can be instead attributed to a magnon. This conclusively demonstrates that individual magnons can carry electric dipole moment in \CCMO. Unlike the other cases of magnons showing electric behavior \cite{pimenov2006possible,Pimenov_2008,krivoruchko2012,sirenko2016,pimenov_debye,kida2008,grams2022observation}, the effects we observed persist beyond the ordered phase into the field polarized region $H>H_c$ and are detected at remarkably low temperatures and frequencies.
	
	The obtained value of $10^{-6}$ s for the attempt time $\tau_0$ in \CCMO is striking. It corresponds to a very low attempt frequency of 1~MHz. For most of the field range explored, the magnon frequency is of the order 0.1 THz, i.e. five orders of magnitude higher. This is in stark contrast to other polar insulators \cite{bidault1994,viana1994,ang1999} where $\Delta \tau_0\hbar \sim 0.01 -0.1$. The key diference is that in  \CCMO the low-$T$ dielectric response is likely coming from the lattice, while the relaxation is occurring by the spin excitations. Therefore, $\tau_0$ is determined by the strength of the spin-lattice coupling, which is expected to be much weaker than the phonon interactions in the ferroelectric case.

	In summary, we have demonstrated single-magnon relaxation of dielectric degrees of freedom in \CCMO.  The applied magnetic field becomes a convenient ``knob'' for tuning interactions between the polarizeable lattice and spins due to proximity to a magnetic quantum critical point. These results open a new route to  multiferroic quantum criticality
    \cite{Narayan2019} and potentially  to coupling magnetic, electric 
    and optical degrees of 
	freedom in hybrid quantum circuits \cite{lachance2019hybrid,clerk2020hybrid}.

 This work was partially supported by the Swiss National Science Foundation, Division II.
	P.C. is funded by DOE Basic Energy Sciences grant DE-SC0020353 and P.A.V. was supported by a  Rutgers Center for Materials Theory Abrahams Fellowship when this project was initiated.
	P.C. and P.A.V. acknowledge the Aspen Center for Physics where this work was discussed, which is supported by National Science Foundation grant PHY-1607611 and a grant from the Simons Foundation (P.A.V.).

\bibliography{Main_prl.bib}
\end{document}


\title{Supplemental material to: Dielectric relaxation by quantum critical magnons}	
	
	\author{Daniel Flavi\'{a}n}
	\affiliation{Laboratory for Solid State Physics, ETH Z{\"u}rich, 8093 Z{\"u}rich, Switzerland}
	\author{Pavel A. Volkov}
	\affiliation{Department of Physics, Harvard University, Cambridge, Massachusetts 02138, USA}
	\affiliation{Department of Physics, University of Connecticut, Storrs, Connecticut 06269, USA}
	\affiliation{Department of Physics and Astronomy, Center for Materials Theory, Rutgers University, Piscataway, NJ 08854, USA}
	\author{Shohei Hayashida}
	\affiliation{Laboratory for Solid State Physics, ETH Z{\"u}rich, 8093 Z{\"u}rich, Switzerland}
	\affiliation{Present address: Max-Planck-Institut f\"ur Festk\"orperforschung, Heisenbergstraße 1, 70569 Stuttgart, Germany}
\author{K.~Yu.~Povarov}
	\affiliation{Laboratory for Solid State Physics, ETH Z{\"u}rich, 8093 Z{\"u}rich, Switzerland}
	\affiliation{Present address: Dresden High Magnetic Field Laboratory (HLD-EMFL) and W\"urzburg-Dresden Cluster of Excellence ct.qmat, Helmholtz-Zentrum Dresden-Rossendorf, 01328 Dresden, Germany}	
	\author{Severian Gvasaliya}
	\affiliation{Laboratory for Solid State Physics, ETH Z{\"u}rich, 8093 Z{\"u}rich, Switzerland}
	\author{Premala Chandra}
	\affiliation{Department of Physics and Astronomy, Center for Materials Theory, Rutgers University, Piscataway, NJ 08854, USA}
	\author{Andrey Zheludev}
	\affiliation{Laboratory for Solid State Physics, ETH Z{\"u}rich, 8093 Z{\"u}rich, Switzerland}
	
\maketitle

\section{Experimental details}

\CCMO crystallizes in a monoclinic structure (space group $C2/c$). 
Its magnetic properties stem from highly-frustrated exchange interactions along the crystallographic \textit{b} axis, where Cu$^{2+}$ ions run in chains. A detailed investigation of the chemical structure of this material is still lacking in the literature. Therefore, we performed a meticulous study of the structure as shown below. 

All of the complex permittivity measurements reported in the main text have been carried out on single crystal samples. The crystal growth and the thermodynamic properties are detailed somewhere else \cite{Flavian_2020}. Silver paste electrodes were deposited on opposing faces of a crystal of dimensions 2.4x0.24x0.1 mm$^3$ to build a capacitor. The sample was then installed on a sapphire substrate to ensure a good thermal exchange with the heat bath at low temperatures. 

The measurements were carried out using a $^3$He-$^4$He dilution refrigerator insert for the Quantum Design Physical Property Measurement System (PPMS). Measurements of dielectric capacitance were performed with two different setups. Data at 1 kHz were measured on a Andeen-Hagerling AH 2550A Ultra-precision Capacitance Bridge (optimized for measurements at 1kHz). Measurements as a function of probing frequency were carried out in the Andeen-Hagerling AH 2700A Ultra-precision Capacitance Bridge. Due to the small amplitude of the measured signals, the data were taken always at constant frequencies as the temperature and magnetic field were modified continuously. Typically, a rate of 0.01 K/min was chosen for the temperature scans. For field scans the ramping rate was optimized in order to minimize the spurious heating due to eddy currents. A rate of 2.5 Oe/s was chosen below 0.25 K, 5 Oe/s between 0.25 K and 0.40 K and 10 Oe/s for all measurements above 0.40 K.  In order to maximize the signal to noise ratio while keeping the self-heating minimal an excitation signal of 0.75 V was used in the measurements.

\section{Crystal structure refinement}

\begin{figure}[tbp]
    \centering
    \includegraphics[scale=1]{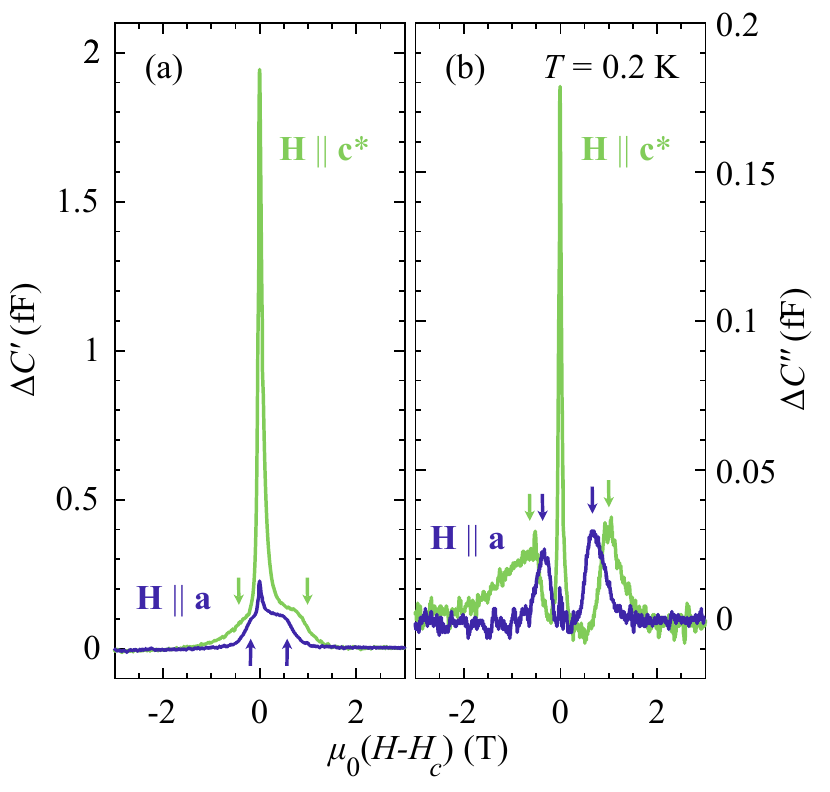}
    \caption{Comparison of the dielectric signal obtained for magnetic fields applied along orthogonal crystal directions \textbf{H}$\parallel$\textbf{a} and \textbf{H}$\parallel$\textbf{c}$^\ast$.  To ease the comparison, the fields are plotted with respect to saturation fields. Both the (a) real and (b) imaginary components show the large difference in the peaks at the saturation, while the dissipative feature (arrows) is hardly affected by  geometry. }
    \label{fig:Compare_axes}
\end{figure}

The crystal structure of \CCMO has been discussed in the literature by several authors \cite{hase2005magnetism,fujimura2016comparison,hamasaki2007effects, hoshino2014ground,goto2017nmr,yagi2018133cs,matsui2018ground}. In the original paper \cite{solodovnikov1997new},  no determination of the atomic coordinates is provided.  A thorough refinement of the chemical structure was still lacking. Here we provide such a refinement of single crystals at 300 K and at 100 K. Our results confirm the expected crystal structure at both temperatures and show no change in symmetry down to 100 K. 

The structure was solved using a prismatic crystal with approximate dimensions 30x10x10 $\mu$m$^3$ in a Bruker APEX-II instrument. Preliminary cell parameters were determined. A subsequent data collection was carried out to refine the crystal parameters and identify the chemical structure. An elevated number for redundancy of the measured reflections was selected to allow for an automated correction for X-ray absorption. Reflection intensities were analyzed and transformed into structural amplitudes using the suites SHELXL-2014. 

After a few rounds of refinement the structure converged to a space group $C2/c$ with the correct chemical composition. Some of the oxygen ions needed to be fixed in place manually to improve convergence. The final structure yielded a good agreement $R = 5.6 $\%. The inclusion of anisotropic thermal parameters for all atoms give a final refinement of $R = 4.3$\%. The fractional coordinates for all atoms from the refinement are summarized in table \ref{tab:Refinement_300_positions}, alongside with the equivalent isotropic thermal parameters. Note that, having a monoclinic structure, these are calculated as 
\begin{equation}
    U_{eq} = \left( U_{11} + U_{22} \sin^2{\beta} + U_{33} + 2U_{23}\cos{\beta} \right)/3\sin{\beta}
    \label{Eq:U_eq}
\end{equation}
A refinement of the structure at $T=$ 100 K was carried out on the same sample. The high temperature structure was taken as a starting point for the refinement of the low temperature data. A good refinement was readily obtained, with $R=5.4$\%.  Further inclusion of anisotropic thermal factors yields a final agreement factor of $R=3.1$\%. 

\begin{figure}[tbp]
    \centering
    \includegraphics[scale=1]{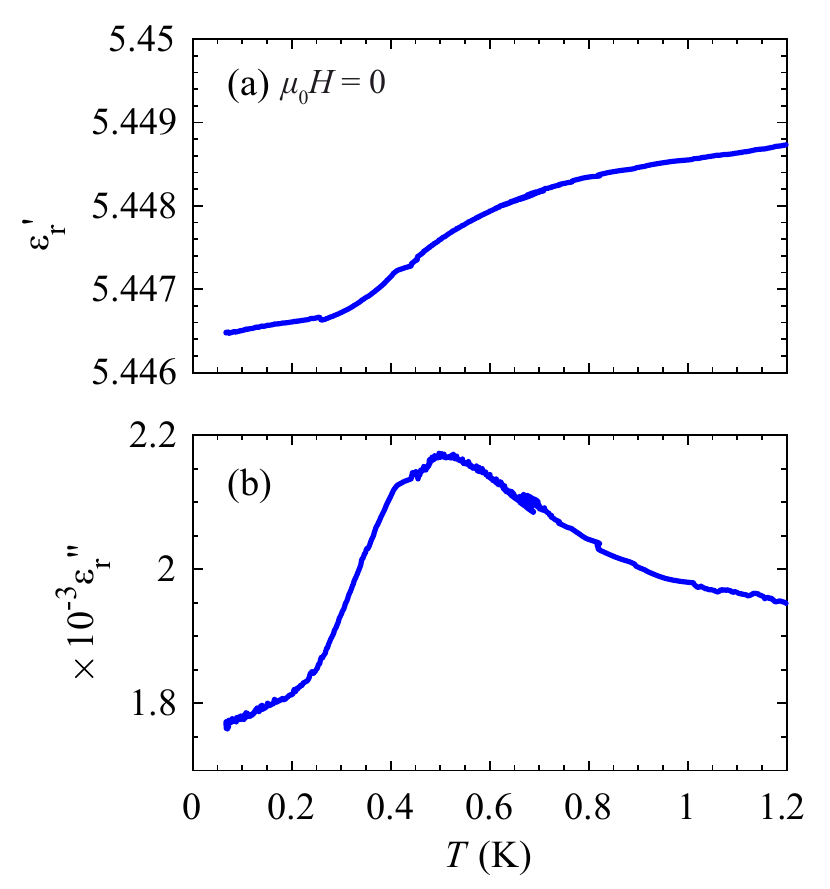}
    \caption{Relative complex permittivity ((a) real, and (b) imaginary components) of a polycrystalline sample of \CCMO at low temperatures.  }
    \label{fig:Relative_epsilon}
\end{figure}

Our refinements are in agreement with the structure reported in \cite{solodovnikov1997new}. Importantly, they show that the complex structure can correctly be described in a $C2/c$ space group and that the structure remains unchanged down to 100 K. The refined space group is centrosymmetric. The presence of inversion center therefore forbids the spontaneous appearance of a ferroelectic dipole moment.  Cooling down from 300 K to 100 K is mainly reflected in a shortening of the crystallographic \textit{a} axis, while the other axes remain almost unchanged. 

\begin{table*}
  \caption{Summary of X-ray refinements of the chemical structure of \CCMO at 100 K and 300 K.}
  \centering 
  \begin{threeparttable}
    \begin{tabular}{c| c| c}
    Characteristics  & $T$ = 100 K  & $T$ = 300 K\\
     \midrule\midrule
     Crystal system & \hspace{1.5cm}Monoclinic\hspace{1cm} & \hspace{1cm}Monoclinic\hspace{1cm}  \\  
     \cmidrule(l  r ){1-3}
     
     Space group & $C2/c$ & $C2/c$ \\ 
     \cmidrule(l  r ){1-3}
     
     \textit{a}, \AA &  27.625(6) & 27.9224(18)  \\
     \cmidrule(l  r ){1-3}
     
     \textit{b}, \AA & 5.121(2) &  5.1265(3)  \\
     \cmidrule(l  r ){1-3}
     
     \textit{c}, \AA & 20.123(4) & 20.2592(13)  \\
     \cmidrule(l  r ){1-3}
     
     $\beta$, deg & 106.88(3) & 107.280(2)  \\
     \cmidrule(l  r ){1-3}
     
     V, \AA$^3$ & 2724.1(14) &  2769.1(3)  \\  
     \cmidrule(l  r ){1-3}
     
     Z & 8 & 8 \\ 
     \cmidrule(l  r ){1-3}
     
     $d_{calc}$, g/cm$^3$ & 4.23 & 4.18 \\
     \cmidrule(l  r ){1-3}
     
     2$\theta_{max}$, deg & 56.082 & 62.822 \\
     \cmidrule(l  r ){1-3}
     
     Measured reflections & 44340 & 68466  \\
     \cmidrule(l  r ){1-3}
     
     Independent reflections & 3304 & 4578  \\ 
     \cmidrule(l  r ){1-3}

     Refined parameters & 174 & 174   \\
     \cmidrule(l  r ){1-3}
     
     R & 0.031 & 0.043   \\

    \midrule\midrule
    \end{tabular}
    
    \end{threeparttable}
    \label{tab:Refinement}
\end{table*}

\begin{table*}
  \caption{Fractional coordinates and equivalent isotropic thermal parameters of the atomic basis of \CCMO at 100 K.}
  \centering 
  \begin{threeparttable}
    \begin{tabular}{c|c c c|c}
    \hspace{1cm}Atom\hspace{1cm}  & \hspace{1cm}$x/a$\hspace{1cm}  & \hspace{1cm}$y/b$\hspace{1cm} & \hspace{1cm}$z/c$\hspace{1cm} & \hspace{1cm}$U_{eq}$\hspace{1cm} \\
     \midrule\midrule

        Cs(1) & 0 & 0.17359(11) & 0.75 & 0.01222(14)\\ 
     \cmidrule(l  r ){1-5}
     
        Cs(2) & 0.32191(2) & 0.18638(8) & 0.23093(2) & 0.01217(12)\\ 
     \cmidrule(l  r ){1-5}
     
        Cs(3) & 0.25 & 0.25 & 0.5 & 0.01542(15)\\ 
     \cmidrule(l  r ){1-5}
     
        Mo(1) & 0.01712(2) & 0.23496(9) & 0.06894(3) & 0.00728(14)\\ 
     \cmidrule(l  r ){1-5}
     
       	Mo(2) & 0.18263(2) & 0.25394(9) & 0.10717(3) & 0.00780(14)\\ 
     \cmidrule(l  r ){1-5}
     
        Mo(3) & 0.11275(2) & 0.25156(9) & 0.36754(3) & 0.00736(14)\\ 
     \cmidrule(l  r ){1-5}
     
        Cu(1) & 0.42254(3) & 0.24693(13) & 0.04626(4) & 0.00758(17)\\ 
     \cmidrule(l  r ){1-5}
     
        Cu(2) & 0.40071(3) & 0.256883(13) & 0.04204(4) & 0.00767(18)\\ 
     \cmidrule(l  r ){1-5}

        O(1) & 0.5562(16) & 0.4806(9) & 0.6241(2) & 0.0147(9)\\ 
     \cmidrule(l  r ){1-5}
          
        O(2) & 0.4972(15) & 0.0545(9) & 0.6213(2) & 0.0128(9)\\ 
     \cmidrule(l  r ){1-5}
          
        O(3) & 0.44633(14) & 0.9437(8) & 0.4757(2) & 0.0090(8)\\ 
     \cmidrule(l  r ){1-5}
     
        O(4) & 0.4586(14) & 0.4494(8) & 0.5188(2) & 0.0090(8)\\ 
     \cmidrule(l  r ){1-5}
     
        O(5) & 0.30031(17) & 0.6361(9) & 0.3084(2) & 0.0139(9)\\ 
     \cmidrule(l  r ){1-5}
     
        O(6) & 0.26170(17) & 0.7837(8) & 0.4161(2) & 0.0145(10)\\ 
     \cmidrule(l  r ){1-5}

        O(7) & 0.34640(16) & 0.0640(8) & 0.3916(2) & 0.0140(9)\\ 
     \cmidrule(l  r ){1-5}
     
        O(8) & 0.35953(15) & 0.5446(8) & 0.4561(2) & 0.0118(8)\\ 
     \cmidrule(l  r ){1-5}
     
        O(9) & 0.4218(15) & 0.1071(9) & 0.7905(2) & 0.0119(8)\\ 
     \cmidrule(l  r ){1-5}
    
        O(10) & 0.3266(17) & 0.2921(8) & 0.6389(3) & 0.0123(9)\\ 
     \cmidrule(l  r ){1-5}
          
        O(11) & 0.4141(16) & 0.5632(8) & 0.6247(2) & 0.0129(9)\\ 
     \cmidrule(l  r ){1-5}
     
        O(12) & 0.3836(15) & 0.0487(8) & 0.5567(2) & 0.0097(8)\\ 
    \midrule\midrule
    \end{tabular}
    
    \end{threeparttable}
    \label{tab:Refinement_100_positions}
\end{table*}

\begin{table*}
  \caption{Fractional coordinates and equivalent isotropic thermal parameters of the atomic basis of \CCMO at 300 K.}
  \centering 
  \begin{threeparttable}
    \begin{tabular}{c|c c c|c}
    \hspace{1cm}Atom\hspace{1cm}  & \hspace{1cm}$x/a$\hspace{1cm}  & \hspace{1cm}$y/b$\hspace{1cm} & \hspace{1cm}$z/c$\hspace{1cm} & \hspace{1cm}$U_{eq}$\hspace{1cm} \\
     \midrule\midrule

        Cs(1) & 0 & 0.17028(9) & 0.75 & 0.02962(12)\\ 
     \cmidrule(l  r ){1-5}
     
        Cs(2) & 0.32211(2) & 0.18822(7) & 0.22993(2) & 0.03142(10)\\ 
     \cmidrule(l  r ){1-5}
     
        Cs(3) & 0.25 & 0.25 & 0.5 & 0.04041(15)\\ 
     \cmidrule(l  r ){1-5}
     
        Mo(1) & 0.01705(2) & 0.24338(7) & 0.06900(2) & 0.01479(9)\\ 
     \cmidrule(l  r ){1-5}
     
        Mo(2) & 0.18110(2) & 0.24739(7) & 0.10722(2) & 0.01687(10)\\ 
     \cmidrule(l  r ){1-5}
     
        Mo(3) & 0.11195(2) & 0.24560(6) & 0.36868(2) & 0.01471(9)\\ 
     \cmidrule(l  r ){1-5}
     
        Cu(1) & 0.42334(2) & 0.25615(9) & 0.04620(3) & 0.01479(12)\\ 
     \cmidrule(l  r ){1-5}
     
        Cu(2) & 0.40146(2) & 0.24091(9) & 0.04231(3) & 0.01536(12)\\ 
     \cmidrule(l  r ){1-5}

  
        O(1) & 0.55577(13) & 0.4778(7) & 0.62348(18) & 0.0269(7)\\ 
     \cmidrule(l  r ){1-5}
     
        O(2) & 0.49819(13) & 0.0508(7) & 0.62106(18) & 0.0271(7)\\ 
     \cmidrule(l  r ){1-5}
     
        O(3) & 0.44687(11) & 0.9397(6) & 0.47625(16) & 0.0170(6)\\ 
     \cmidrule(l  r ){1-5}
     
        O(4) & 0.45907(11) & 0.4475(6) & 0.51933(16) & 0.0170(6)\\ 
     \cmidrule(l  r ){1-5}
     
        O(5) & 0.30076(15) & 0.6362(7) & 0.30905(18) & 0.0314(8)\\ 
     \cmidrule(l  r ){1-5}
     
        O(6) & 0.26433(16) & 0.7868(8) & 0.4158(2) & 0.0367(10)\\ 
     \cmidrule(l  r ){1-5}
     
        O(7) & 0.34876(13) & 0.0594(6) & 0.39121(18) & 0.0284(8)\\ 
     \cmidrule(l  r ){1-5}
     
        O(8) & 0.36040(11) & 0.5398(6) & 0.45536(17) & 0.0213(6)\\ 
     \cmidrule(l  r ){1-5}
     
        O(9) & 0.42211(13) & 0.1022(7) & 0.70804(18) & 0.0280(7)\\ 
     \cmidrule(l  r ){1-5}
   
        O(10) & 0.32795(13) & 0.2901(7) & 0.6372(2) & 0.0266(7)\\ 
     \cmidrule(l  r ){1-5}
     
        O(11) & 0.41481(13) & 0.5589(6) & 0.62353(16) & 0.0239(7)\\ 
     \cmidrule(l  r ){1-5}
     
        O(12) & 0.38478(11) & 0.0439(6) & 0.55657(16) & 0.0179(6)\\ 
		
    \midrule\midrule
    \end{tabular}
    \end{threeparttable}
    \label{tab:Refinement_300_positions}
\end{table*}

\section{Comparison of different axes}

The figures in the main text show how the relaxing features observed in \CCMO are virtually isotropic and independent on the direction of the externally applied magnetic field. Here we show explicitly how the divergent peak at the quantum critical point compares to the relaxing features and the effect of the direction the applied external field. 

In Fig.~\ref{fig:Compare_axes} the capacitance measured at a constant temperature of 0.2 K for two orthogonal fields is compared.  It is evident from the figure the anisotropic response at the saturation transition. The measured signal is roughly ten times larger when the external field is applied along  \textbf{c}$^\ast$. In contrast, the amplitude of the relaxing signal (arrows) remains equal for both data sets. The apparent larger amplitude above $H_c$ for \textbf{H}$\parallel$\textbf{c}$^\ast$ is likely due to the tail of the divergent feature. 

\section{Estimation of the relative permittivity}

In order to characterize the absolute complex dielectric permittivity,  we performed measurements on a large powder sample where we had an optimal control of the sample dimensions.  A cylindrical pellet-like sample of 5.16 mm in diameter and 0.66 mm in width was used to measure the data displayed in Fig. ~\ref{fig:Relative_epsilon}. The powder was compressed using a hydrostatic pressure of 80 kbar and subsequently sintered for 24 h at 500 $^\circ$C.  Silver electrodes were deposited on it using the same procedure as for the single crystal samples. 

Data in Fig.~\ref{fig:Relative_epsilon} reveal a modest relative dielectric constant of $\epsilon_r = $ 5.44. This value is similar to that reported for the Rb-based sister compound \cite{Reynolds_2019,Ueda_2020}. Note, however, that all the dielectric studies of \RCMO reported values of $\epsilon$ on samples containing ferroelectric impurities \cite{Hayashida2021}, which contributed to an overestimation of the dielectric constant of \RCMO. 

Notably, the data in Fig.~\ref{fig:Relative_epsilon} show the same features discussed in the main text: a peak in $C^{\prime\prime}$ around 0.45 K, and a step in capacitance at the same temperature. While this feature is possibly broadened by the powder-nature of the sample, it strongly suggest that relaxation is not happening at the surface of the material but is rather an intrinsic bulk effect of \CCMO.  

\section{Extrinsic resonances}

\begin{figure}[tbp]
    \centering
    \includegraphics[scale=1]{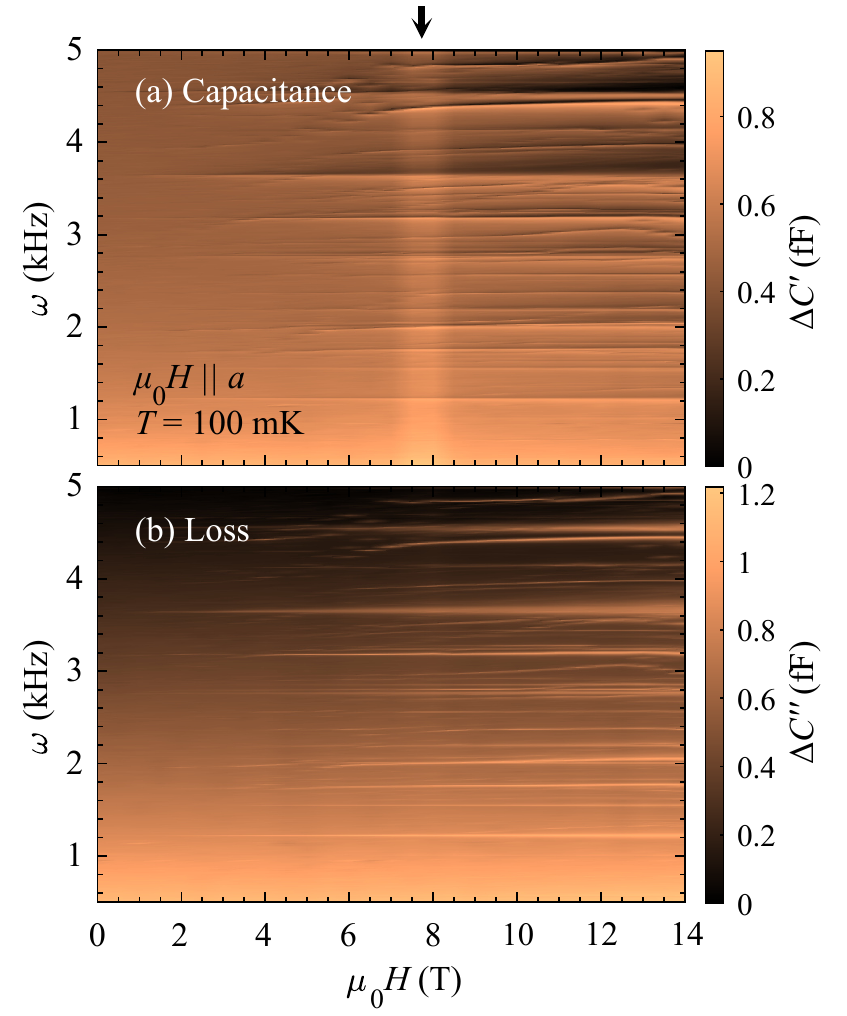}
    \caption{Complex capacitance scans at 100 mK at different magnetic fields and frequencies: (a) real component, (b) imaginary component. Resonant features can be identified at a variety of frequencies, whose strength grows as does the magnetic field (\textbf{H}$\parallel$\textbf{a}). In (a), a faint vertical stripe represents the peak at the critical field (black arrow), which gives an idea of the relative strength of the resonances with respect to the signal measured from the sample.}
    \label{fig:Resonances}
\end{figure}

A systematic study of the response of the system under different excitation frequencies was carried out. To single out the origin of the anomalies found at 1 kHz, a wide range of frequencies was tested. Measurements at constant temperature of 100 mK were carried out between 0.5 and 5 kHz for the full range of magnetic fields. The results are displayed in Fig.\ref{fig:Resonances} and show a complex landscape of anomalies. All of these anomalies correspond to electro-mechanical resonances in our measurement setup. As such, the real and imaginary part of the response function (here $\Delta C^{\prime}$ and $\Delta C^{\prime\prime}$, respectively) must be related via Kramers-Kronig relations. This fact was checked for the most prominent features at a variety of fields, confirming the resonant nature. The amplitude of these features grows almost linearly proportional to the strength of the magnetic field, pointing to a linear coupling to it. 

These modes are rather strong in comparison with the properties that are discussed in the main text. The anomaly corresponding to the saturation field ($\mu_0H_\textit{a} = 7.66$ T) is visible as a vertical stripe in the data set in Fig.\ref{fig:Resonances}.a. From this we obtain an amplitude ratio of up to 10:1 between the additional resonances and the saturation transition peak. 

Two observations from the pattern of resonances are to be noted. First, the resonance frequency is essentially field and temperature independent for the majority of observed features.  Secondly, there is no normal modes below 1.2 kHz and, in addition, the lowest frequency mode is narrow in frequencies all the way up to 14 T. This in turn shows that our measurements at 1 kHz are protected from these mechanical resonances. 

We are able to understand these resonances in terms of a simple toy model. The system under study can be simplistically thought of as composed of a pair of capacitor plates held at the cold finger of a refrigerator. In order to measure capacitance, long, high-fidelity coaxial cables connect the sample to the measurement device, though leaving a small exposed piece of cabling that creates effectively a coil, exposed to the magnetic field.  The cold finger is subject to vibrations that  move the coil in the field, generating in it a small electric current by induction. This effect is enhanced when the probing frequency of the electric field matches one of the mechanical normal modes of the cold finger system, giving rise to resonant behavior via Lorentz forces.  Consequently, these modes are completely extrinsic and unrelated to the physics of the system under study. They can, however, not be avoided either, as they are a result of the geometry of the experiment. For this reason, measuring magnitudes against frequency is not advisable in our experiment, as the largest contributions come from resonant features, masking the subtle features under study.  Notably, however, the resonances are absent in the absence of an external field. Therefore, our zero-field measurements are unaffected by extrinsic resonances.  

\section{Determination of attempt time}

\begin{figure}[tbp]
    \centering
    \includegraphics[scale=1]{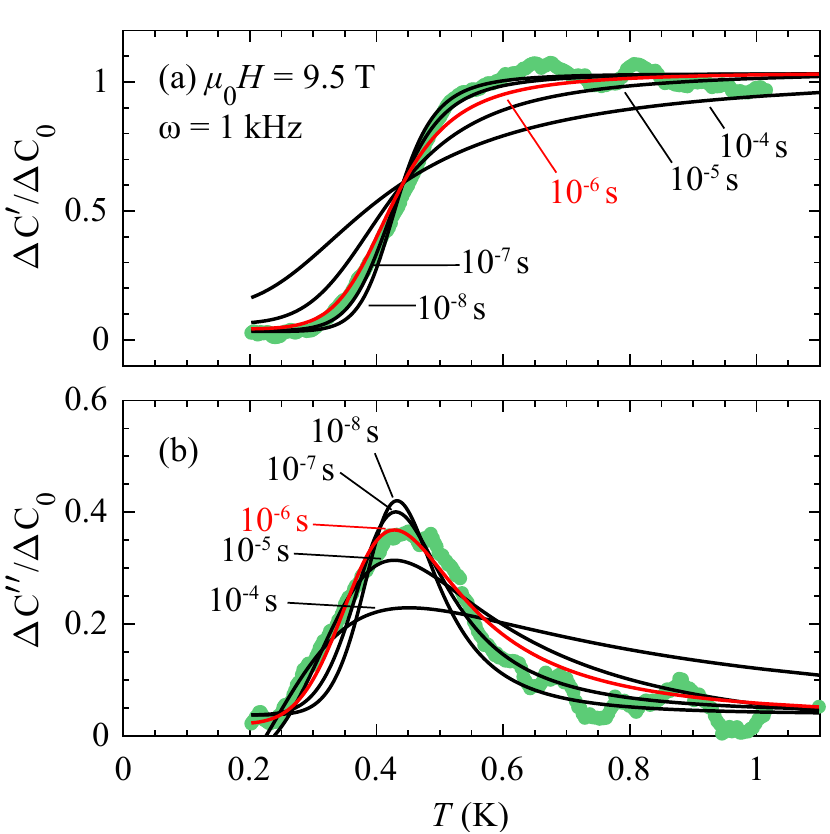}
    \caption{The goodness of fit to \eqref{eq:ColeCole} strongly depends on the attempt time, $\tau_0$. The panels show the (a) real and (b) imaginary components of an instance of the best simultaneous fits with a fixed attempt time, displayed next to the curves. The best fit with $\tau_0= 10^{-6} s$ is highlighted in red.}
    \label{fig:Tau_0}
\end{figure}

The Cole-Cole model of equation \eqref{eq:ColeCole} with an activated behavior shows a highly nonlinear dependence on several parameters. To avoid overparametrization we assume a single 'attempt time', $\tau_0$, that is field and temperature independent. The value of $\tau_0$ is optimized using the whole set of data presented in the main text and then fixed for a global fit. In Fig.~\ref{fig:Tau_0} the effect of the choice of different values of $\tau_0$ is exemplified. We run a global fit of \eqref{eq:ColeCole} to our data under different assumptions for $\tau_0$ and the results are displayed in the figure. A value of $\tau_0\approx10^{-6}$~s  shows overall the best agreement with the experimental data.

\section{Cole-Cole plots}

\begin{figure}[tbp]
    \centering
    \includegraphics[scale=1]{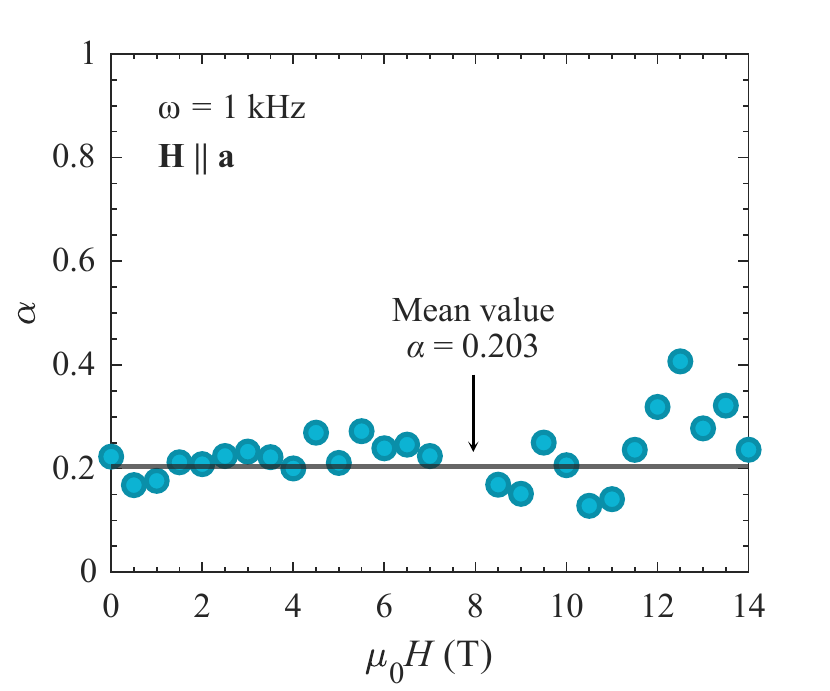}
    \caption{Phenomenological $\alpha$ parameter extracted from the best fit of data in Fig.~\ref{fig:ColeCole} to \eqref{eq:ColeCole}. The average value, used in the global fit in the main text is shown as a solid line.}
    \label{fig:AlphaValues}
\end{figure}

\begin{figure*}[tbp]
    \centering
    \includegraphics[scale=1]{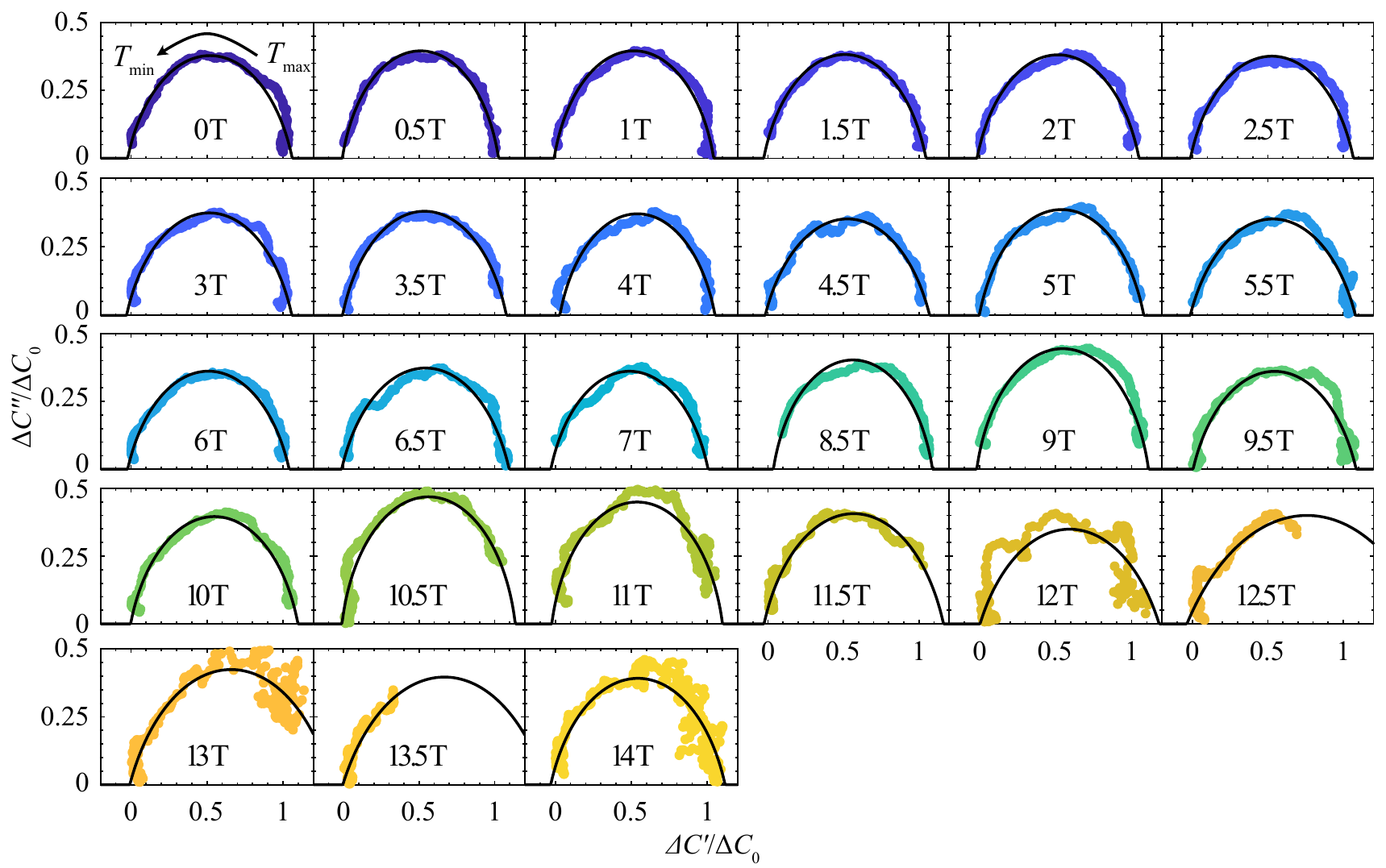}
    \caption{Cole-Cole plots for data collected with \textbf{H}$\parallel$\textbf{a} with a probing frequency of 1 kHz. Axes are shared across all subplots. Black solid lines represent the best fit to a semi-circumference as expected for the Cole-Cole model. The extracted values for $\alpha$ are summarized in \ref{fig:AlphaValues}. }
    \label{fig:ColeCole}
\end{figure*}

In real materials the existence of imperfections (such as impurities, dislocations, pinning centers, etc.) leads to a deviation from the ideal Debye-Model for relaxation. Generally, a certain distribution of relaxation times appears that can be modeled by phenomenological parameters. As shown in the main text, the studied effects can be well captured in terms of Cole-Cole relaxation, where a parameter $\alpha$ is introduced as follows: 
\begin{equation}
    \Delta \varepsilon  = \frac{\varepsilon_0(T)-\varepsilon_\infty}{1+(i\omega\tau(T))^\alpha}
    \label{eq:ColeCole}
\end{equation}
Under these conditions a plot of $\varepsilon^{\prime}$ vs. $\varepsilon^{\prime\prime}$ lies on a circumference. Since its center is displaced out of the abscissa axis by a factor proportional to $1-\alpha$, the Cole-Cole plots shown in Fig.\ref{fig:ColeCole} are a perfect tool to unequivocally identify this parameter. It is customary plot such data as a function of the probing frequency $\omega$. Here, however, we collected the bulk of our data varying the temperature at a constant frequency of 1 kHz. The data in Fig.\ref{fig:ColeCole} is plotted as a function of temperature. With increasing temperature, the system transits between the low temperature regime (where $\varepsilon=\varepsilon_\infty$, equivalent to the high frequency regime) and the high temperature regime (where $\varepsilon=\varepsilon_{static}$, low frequency regime). Since the relevant quantity in the model is the product $\omega\tau$, the global effect must be independent of what parameter (either $\omega$ or $\tau$) is varied.

The data in Fig.\ref{fig:ColeCole} are fitted to the Cole-Cole model at a constant frequency. Independent fits are performed for each value of magnetic field and the results are summarized in Fig.\ref{fig:AlphaValues}. Especially at low fields the data show little variation around the average value of $\alpha_{avg} = 0.20$. This value is then selected to carry out the analysis shown in the main text. The higher discrepancy in the high field data is mainly due to the restricted temperature coverage of the data sets.

\bibliography{Supplement.bib}